\documentclass[pre,twocolumn,aps,showpacs,nofootinbib,superscriptaddress,floatfix]{revtex4}
\newif\ifpdf\ifx\pdfoutput\undefined\pdffalse\else\pdfoutput=1\pdftrue\fi

\usepackage{graphicx}
\usepackage{amsmath}
\newcommand{\beq}{\begin{equation}}
\newcommand{\eeq}{\end{equation}}

\newcommand{\SB}{{\bf S}}

\begin{document}

\title{Function Constrains Network Architecture and Dynamics: A Case Study on the Yeast Cell Cycle Boolean Network} 

\author{Kai-Yeung Lau}
\affiliation{Graduate Group in Biological and Medical Informatics, University of California San Francisco, 1600 16$^{th}$ Street, San Francisco, CA 94158-2517, USA}
\author{Surya Ganguli}
\affiliation{Sloan-Swartz Center for Theoretical Neurobiology, University of California San Francisco,  513 Parnassus Avenue, San Francisco, CA 94143-0444, USA}
\author{Chao Tang}
\altaffiliation{To whom correspondence should be addressed. E-mail: Chao.Tang@ucsf.edu.}
\affiliation{Departments of Biopharmaceutical Sciences and Biochemistry and Biophysics, University of California San Francisco, 1700 4$^{th}$ Street, San Francisco, CA 94143-2540, USA}
\affiliation{Center for Theoretical Biology, Peking University, Beijing 100871, China}
\date{\today}

\begin{abstract}

  We develop a general method to explore how the function performed by
  a biological network can constrain both its structural and dynamical
  network properties.  This approach is orthogonal to prior studies
  which examine the functional consequences of a given structural
  feature, for example a scale free architecture.  A key step is to
  construct an algorithm that allows us to efficiently sample from a
  maximum entropy distribution on the space of boolean dynamical
  networks constrained to perform a specific function, or cascade of
  gene expression.  Such a distribution can act as a ``functional null
  model'' to test the significance of any given network feature, and
  can aid in revealing underlying evolutionary selection pressures on various
  network properties.  Although our methods are general, we illustrate
  them in an analysis of the yeast cell cycle cascade.  This analysis
  uncovers strong constraints on the architecture of the cell cycle
  regulatory network as well as significant selection pressures on this
  network to maintain ordered and convergent dynamics, possibly at the
  expense of sacrificing robustness to structural perturbations.

\end{abstract}


\maketitle

\section{Introduction}


A central problem in biology involves understanding the relationship
between structure and function.  This problem becomes especially
intricate in the era of systems biology in which the objects of study
are biological networks composed of large numbers of interacting
molecules.  To what extent does the structure of a biological network
constrain the range of functions, or types of dynamical behaviors, that
the network is capable of producing?  Conversely, to what extent does the
requirement of carrying out a specific function constrain the structural
and more general dynamical properties of a network?   

There already exists a large body of theoretical work addressing the
former question.  For example Kauffman \cite{Kauffman} and
others performed an extensive study of ensembles of simplified
boolean networks with fixed structural properties, such as the number
of nodes and the mean degree of connectivity.  A principal finding was
a phase transition in the resulting dynamical behavior, from ordered
to chaotic, as the connectivity increased \cite{Ald}.  More recently, the
observation that many biological networks are scale free spurred a
flurry of research into the dynamical consequences of the scale free
structural feature \cite{JeoTomAlbOltBar, AldClu, JeoMasBarOlt,
  MasSne, OikClu}.  A principal finding was that the scale-free architecture
is more robust to random failures and dynamic fluctuations, and may be 
more evolvable.  

Alternatively, the latter question involving the structural and
dynamical consequences of performing a specific function remains
relatively unexplored \cite{Wag05, MaLai, KasAlo}.  It is an important question because many
biological functions are performed by relatively small network modules
for which gross statistical properties such as mean connectivity, or
any kind of degree distribution, scale free or not, do not have a
clear significance.  A key example of such a module is the yeast cell
cycle control network, whose essential function was reduced to the
boolean dynamics of a set of 11 nodes by Li,
et. al. \cite{LiLonLuOuyTan}.  Despite the small network size, a
dynamical analysis of this simple model demonstrated a great deal of
robustness of the cell cycle trajectory to both fluctuations in
protein states and perturbations of network structure.  Is this
robustness carefully selected for through evolution and encoded
somehow in the topological structure of the cell cycle network, or
does it arise for free, simply as a consequence of the functional
constraint of having to produce the long cascade of gene expression
that controls the cell cycle?

In this paper we develop techniques to address this question, and more
generally to address the consequences of specific functional, rather
than structural constraints.  A key step in the above work exploring
the dynamical consequences of fixed structural constraints was the
ability to efficiently sample from the maximum entropy distribution on
the space of networks constrained to have a fixed structural feature,
such as a given degree distribution.  We call such an ensemble 
a {\em structural} ensemble.  Similarly, in order to address
the structural consequences of fixed functional constraints, we
develop an efficient algorithm to sample from a maximal
entropy distribution on the space of biological networks constrained
to perform a specific function.  We call the resulting ensemble of
networks a {\em functional} ensemble.  

Since we use a maximal entropy distribution, which introduces no
further assumptions about the network other than the fact that it
performs a given function, we can use such functional ensembles to
test whether any dynamical or structural property of a given
biological network is simply a consequence of the function it
performs, or rather a consequence of further selection.  Under this
functional null model, statistically significant properties of a
real biological network can give us insight into additional selective
pressures that have operated on that network above and beyond the
baseline functional requirements.   As an illustration of this 
general method, we focus on the specific case of the cell cycle
network of \cite{LiLonLuOuyTan}, uncovering deeper insight into
the evolutionary pressures on its structural and dynamical properties.

\section{Methods}

\subsection{The Dynamical Model}

We consider a simplified, boolean model of biological network
dynamics, in which each degree of freedom, or node $s_i$, $i=1\dots N$
takes one of two values at any given time: either 0 (inactive) or 1
(active).  For example, the two values could signify whether a
protein is expressed or not, or whether a kinase is
activated or not.  Thus the full state at time t is captured by a column vector
\beq
\SB(t) = (s_1(t), s_2(t), \ldots, s_{N}(t))^T
\eeq
that can take one of $2^N$ values.  Time progresses in discrete steps, 
and the nodes can either activate or inhibit each other at the next step.  These
interactions are captured by the network connectivity matrix {\bf C} with elements
$c_{ij}$ representing an interaction arrow from node $j$ to node $i$.
The allowed values of $c_{ij}$ are given by
\beq
c_{ij} \in \lbrack-1,0,1\rbrack.
\eeq
For two (possibly identical) nodes $i$ and $j$, if $c_{ij}$ is nonzero, it can be either activating (+1) or inhibiting (-1).  
This terminology is justfied by the dynamical rule
\beq
s_i(t+1) = f_i({\bf C},\SB(t)),
\label{eq:dynamrule1}
\eeq
where 
\beq
f_i({\bf C},\SB(t)) = \left\{\begin{array}{ll}
                       1,      &\sum_{j} c_{ij}s_j(t) > 0  \\
                       0,      &\sum_{j} c_{ij}s_j(t) < 0  \\
                       s_i(t), &\sum_{j} c_{ij}s_j(t) = 0. 
\end{array}\right.
\label{eq:dynamrule2}
\eeq
Essentially, if the total input to a node is positive (negative), it
will be on (off) at the next time step.  In the case of zero input,
the node maintains its state.

\subsection{Generating Functional Ensembles}

A biological network achieves its function by successfully taking the
values of its nodes through a sequence of states.  Thus we will equate
the notion of function with a specified state trajectory 
\beq \SB(0)
\rightarrow \SB(1) \rightarrow \cdots \rightarrow \SB(T).  
\label{eq:function}
\eeq In our example of the cell cycle network, the state sequence is
simply the natural cell cycle trajectory.  We wish to either enumerate
or uniformly sample from the space of networks, or equivalently
connectivity matrices ${\bf C}$, that can successfully perform the
above sequence of $T$ state transitions.  We can think of each of the
$T$ transitions as providing one constraint on the connectivity matrix
${\bf C}$ via the dynamical rule given by equations
\eqref{eq:dynamrule1} and \eqref{eq:dynamrule2}.  Assuming the nodes
are distinguishable, the number of networks, given by the number of
allowed connectivity matrices is $M \equiv 3^{N^2}$.
Even for small, mesoscopic scale networks such as the cell cycle with
11 nodes, $M \approx 5.39 \times 10^{57}$ and so it is computationally
infeasible to iterate through all of these networks and find those for
which the $T$ constraints corresponding to the $T$ transitions in
\eqref{eq:function} are satisfied.  Even if one were to sample from
these $M$ networks, one would rarely find a network that could perform
the function in \eqref{eq:function}.

However, it is important to note that the constraint on the network
connectivity ${\bf C}$ imposed by a given transition actually
decouples across the rows of the connectivity matrix.  That is for
each node $i$, the dynamical rule in Eq. \eqref{eq:dynamrule2} 
depends only on the $i$'th row $c_{ij}, j=1\dots N$ of ${\bf C}$, or equivalently
on the $N$ incoming interaction arrows whose target is node $i$.  Thus we can check
that the $T$ constraints induced by the target sequence \eqref{eq:function}
are satisfied for each row, independently of the other rows.  For any given
$i$, the number of possible rows is $Z \equiv 3^{N}$, which
is 177,147 for the yeast cell cycle network.  Thus it becomes computationally
feasible to exhaustively iterate through all possible row values, or incoming arrow
combinations, for each node, and check that the $T$ constraints are satisfied
for each such combination.

After following this procedure using the cell-cycle process as the constraint,
let $1 \leq \alpha_i \leq Z$ index the
set of allowed incoming arrow combinations to node $i$ that satisify
all constraints.  If for each node $i$, we uniformly choose a
particular $\alpha_i$, and assemble the corresponding $N$ allowed rows
into a matrix $C$, we will have constructed a network that can
successfully carry out the state trajectory in \eqref{eq:function}.
This is essentially our sampling procedure.  It produces a functional
ensemble: a maximum entropy distribution on the space of networks
constrained to produce the function represented by
\eqref{eq:function}.

\subsection{Combined structural and functional ensembles.}

In order to perform a more fine scale study of the properties of the
yeast cell cycle network, we wish to constrain more than just a
predefined function.  We would also like to understand how various
properties depend on the number of nonzero interaction arrows in the
network.  Thus we need to develop a method to uniformly sample from
the space of networks that both perform a fixed function {\em and} have a
fixed number of arrows.  However, although the algorithm presented
in the previous section uniformly samples the space of networks performing 
a fixed function, when one adds a constraint on the number of arrows,
this algorithm performs a biased sampling of the more constrained space of
networks.  

\par To see the origin of this bias, consider an implementation of the
algorithm.  Let $A$ be the desired number of arrows in the network.
Fix an arbitrary ordering of the nodes $i=1 \dots N$.  Choose node 1 and
pick at random one incoming arrow combination from a uniform
distribution on the allowed incoming arrow combinations to node 1.
Let $a_1$ be the number of nonzero interaction arrows in the chosen
combination.  We then have $A - a_1$ incoming arrows left to
distribute to nodes $2 \dots N$.  If the algorithm continues in this way
all the way to node $N$, one can see that the resulting sampling of
networks is biased in such a way that earlier nodes in the order have
a higher likelihood of receiving more incoming arrows relative to
later nodes.  For example, in the extreme case, the algorithm may run
out of arrows even before it reaches the final node $N$.  To correct for
this bias, at each step of the algorithm corresponding to each node
$i$ we must sample {\em nonuniformly} from a certain conditional 
distribution on the allowed incoming arrow combinations to node $i$. 

We perform this nonuniform sampling as follows.  Suppose that by the
time we have reached node $i$, we have $r_i$ incoming arrows left to
distribute among the remaining nodes from $i \dots N$.  By definition,
$r_1 = A$, the number of desired arrows in our ensemble.  To choose a
particular incoming arrow combination to node $i$, we first randomly
draw the number of incoming arrows $a_i$ we wish to assign to node $i$
from a conditional probability distribution $P_i(a_i|r_i)$.  Then we
draw the particular combination from a uniform distribution on the set
of allowed incoming arrow combinations that have exactly $a_i$ nonzero
arrows.  $P(a_i|r_i)$ is a conditional distribution on the number of
nonzero incoming arrows to node $i$, conditioned on the total number
of incoming arrows to the remaining unassigned nodes $i\dots N$, and
it can be computed as follows.  Let $W_i(a_i)$ be the number of
allowed incoming arrow combinations to node $i$ that have exactly
$a_i$ nonzero arrows.  Let $Q_{i+1}(b)$ be the number of ways to
distribute $b$ incoming arrows (consistent with the fixed function the
network must perform) to the remaining nodes $i+1 \dots N$.  Then for
each choice $a_i$ of the number of incoming arrows to node $i$, the
quantity $W_i(a_i)Q_{i+1}(r_i-a_i)$ is the number of allowed ways to
complete the network.  $P(a_i|r_i)$ is simply proportional to this
number:
 \beq P_i(a_i|r_i)=
\frac{W_i(a_i)Q_{i+1}(r_i-a_i)}{\sum_{b=0}^{min(N,r_i)}W_i(b)Q_{i+1}(r_i
  - b)}, \quad i=1 \dots N-1.
\label{eq:pai}
\eeq 
Intuitively, the combinatoric factor $Q_{i+1}(r_i-a_i)$ in the
definition of of the conditional distribution $P_i(a_i|r_i)$ corrects
for the biased sampling mentioned above by forcing the algorithm to
choose uniformly from the set of possible completions of the network,
rather than simply uniformly from the set of allowed incoming arrows
to node $i$.  For the special case of the last node $i=N$, when we
have $r_N$ arrows left to distribute, we simply choose uniformly from
the allowed space of incoming arrow combinations to node $N$ that have
exactly $r_N$ nonzero arrows.  Furthermore, each time we run the
algorithm to obtain a network, we randomize the ordering on the nodes.

\par
For any fixed ordering of the nodes, the combinatoric factors $Q_i(b)$
are crucial to the success of the sampling algorithm.  We note that
these factors can be computed efficiently by working recursively from
node $i=N$ back down to $i=1$.  More precisely, if we define 
\beq
Q_N(b) \equiv W_N(b),
\eeq
then for each $i = 1 \dots N-1$ we have the recursion relation
\beq
Q_i(b) = \sum_{a=0}^{min(N,b)} W_i(a)Q_{i+1}(b-a).
\eeq

Thus this algorithm gives us an efficient way to sample from a combined 
structural and functional ensemble: a maximum entropy distribution on the
space of networks constrained to have a fixed function and a fixed number
of arrows.  It is this algorithm that we use to generate the ensemble of ``cell cycle''
networks described below.  In order to compare this ensemble to a set
of more random networks that serve no particular function, we
generated this ``random network'' ensemble by randomly rewiring the
connections in each ``cell-cycle'' network under the constraint that
all nodes must be connected to the same network, {\it i.e.} no
isolated nodes or sub-networks.

\begin{figure}
\includegraphics[width=\columnwidth]{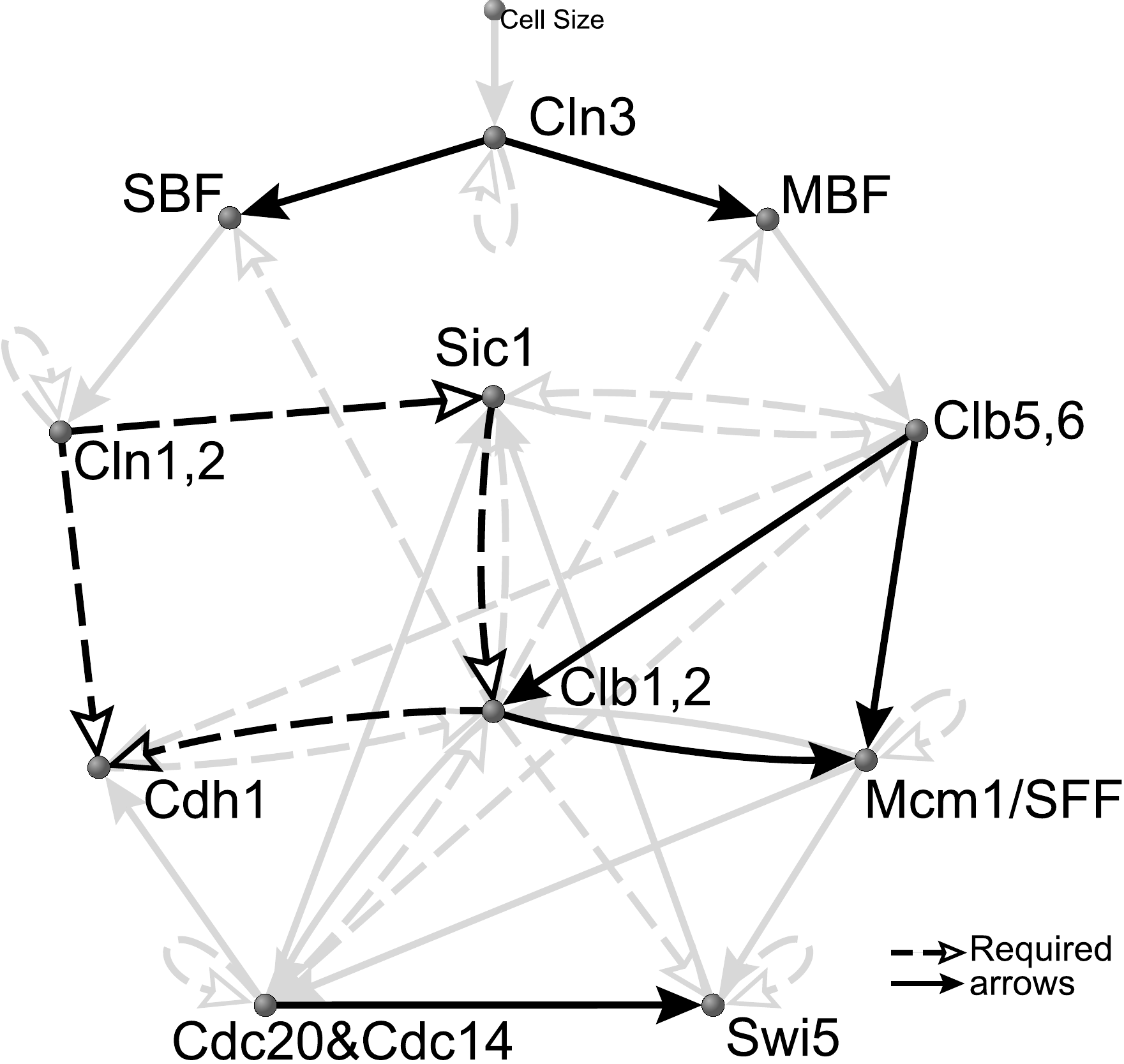}
\caption
{Simplified yeast cell-cycle network from Li {et al.}~\cite{LiLonLuOuyTan}.
Each node represents a gene or its protein product that participates in the regulation of the yeast cell-cycle.
There are cyclins (Cln1,2, Cln3, Clb1,2, Clb5,6), transcription factors (SBF, MBF, Mcm1/SFF, Swi5), and
factors that inhibit or degrade cyclins (Sic1, Cdh1, Cdc20).
Solid and dashed arrows are activating and deactivating interactions, respectively. 
The black solid and black dashed arrows are absolutely required for a network to produce the cell-cycle process [see Table~\ref{ccprocesstab}].}
\label{networkfig}
\end{figure}
\begin{table*}
\caption
{The cell-cycle process generated by the simplified yeast cell-cycle network in Fig.~\ref{networkfig}.}
\begin{tabular*}{\textwidth}{@{\extracolsep{\fill}} ccccccccccccc}
\hline
\hline
Time  & Cln3 & MBF & SBF & Cln1,2 & Clb5,6 & Clb1,2 & Mcm1 & Cdc20 & Swi5 & Sic1 & Cdh1 & Phase\\
\hline
1 & 1 & 0 & 0 & 0 & 0 & 0 & 0 & 0 & 0 & 1 & 1 & ``Excited'' G$_1$\\
2 & 0 & 1 & 1 & 0 & 0 & 0 & 0 & 0 & 0 & 1 & 1 & G$_1$\\
3 & 0 & 1 & 1 & 1 & 0 & 0 & 0 & 0 & 0 & 1 & 1 & G$_1$\\
4 & 0 & 1 & 1 & 1 & 0 & 0 & 0 & 0 & 0 & 0 & 0 & G$_1$\\
5 & 0 & 1 & 1 & 1 & 1 & 0 & 0 & 0 & 0 & 0 & 0 & S\\
6 & 0 & 1 & 1 & 1 & 1 & 1 & 1 & 0 & 0 & 0 & 0 & G$_2$\\
7 & 0 & 0 & 0 & 1 & 1 & 1 & 1 & 1 & 0 & 0 & 0 & M\\
8 & 0 & 0 & 0 & 0 & 0 & 1 & 1 & 1 & 1 & 0 & 0 & M\\
9 & 0 & 0 & 0 & 0 & 0 & 1 & 1 & 1 & 1 & 1 & 0 & M\\
10 & 0 & 0 & 0 & 0 & 0 & 0 & 1 & 1 & 1 & 1 & 0 & M\\
11 & 0 & 0 & 0 & 0 & 0 & 0 & 0 & 1 & 1 & 1 & 1 & M\\
12 & 0 & 0 & 0 & 0 & 0 & 0 & 0 & 0 & 1 & 1 & 1 & G$_1$\\
13 & 0 & 0 & 0 & 0 & 0 & 0 & 0 & 0 & 0 & 1 & 1 & Stationary G$_1$\\
\hline
\hline
\end{tabular*}
\label{ccprocesstab}
\end{table*}

\section{The yeast cell-cycle network}
The simplified yeast cell-cycle Boolean network [see
Fig.~\ref{networkfig}] given in Li {et al.}~\cite{LiLonLuOuyTan}
contains 11 proteins, or nodes, and 1 checkpoint. There are 34 arrows
connecting the nodes: 15 activating and 19 deactivating (``self-degrading'' 
arrows are equivalent to deactivating arrows under our dynamical model).
Using the above dynamical model, this network can produce the cell-cycle process,
as shown in Table~\ref{ccprocesstab}. Starting from the ``excited'' G$_1$ state,
the process goes through the S phase, the G$_2$ phase, the M phase,
and finally returns to the biological G$_1$ stationary state. The
network also has 7 fix-points, with the G$_1$ stationary state being
the biggest, having a basin size of 1764 ($\approx$ 86\% of all
protein states).

\section{Results}
We used the same 11 nodes and the types of connections in the
simplified yeast cell-cycle Boolean network to construct our ensembles
of networks. Using our technique to generate purely functional ensembles, we were 
able to select
11 sets of inward connection combinations (one for each node) that
produce the cell-cycle process [see
Table~\ref{ccprocesstab}]. Figure~\ref{connectionfig} shows the
compositions of different types of connections in the sets. The number
of selected connection combinations for each node (shown in
parenthesis in Fig.~\ref{connectionfig}) varies for two orders of
magnitude. The number of networks that can realize the cell-cycle
process is $5.11 \times 10^{34}$ and the distribution against the
number of arrows is shown in Fig.~\ref{netcountvsacountfig}. 

\subsection{Constraints on Structure from Function.}
\begin{figure}
\includegraphics[width=\columnwidth]{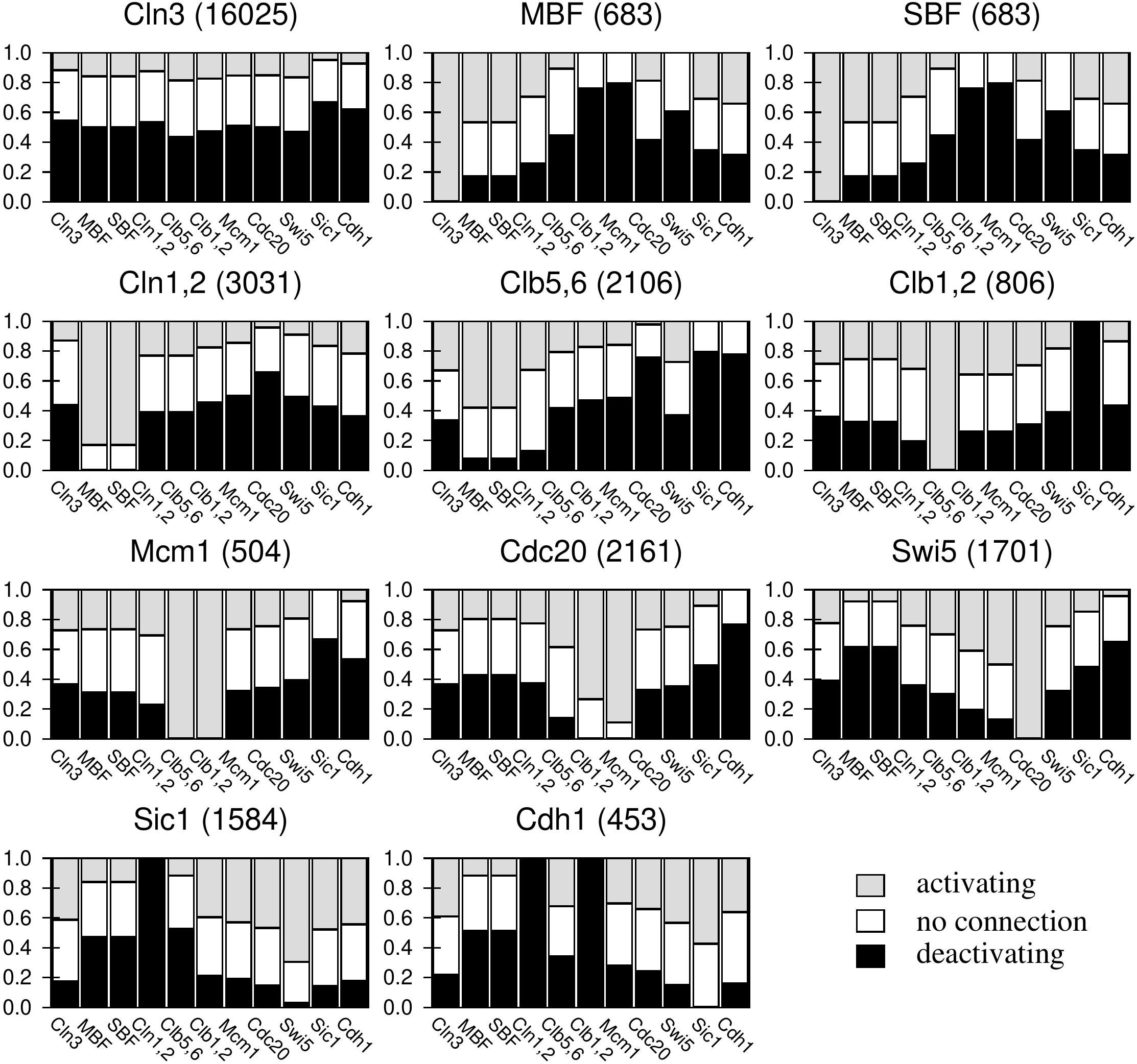}
\caption
{Fractions of different types of inward connections to each node from all other nodes (including itself) that produce the cell-cycle process [see Table~\ref{ccprocesstab}]. Numbers in parenthesis are the number of connection combinations selected by our algorithm [see Methods].}
\label{connectionfig}
\end{figure}

\begin{figure}
\includegraphics[width=\columnwidth]{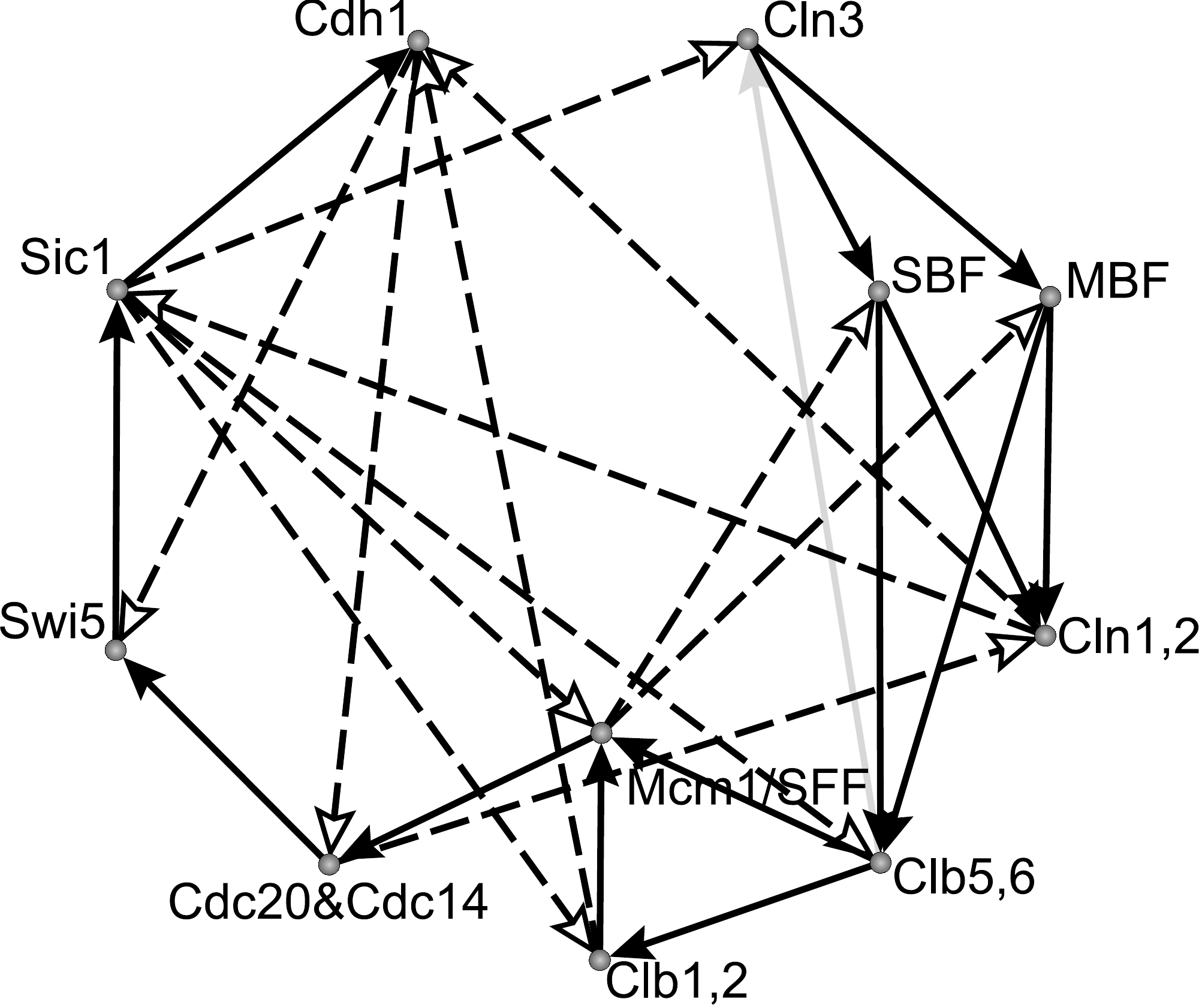}
\caption
{Positive feed-forward and negative feedback connections between network nodes.
Solid and dashed arrows are the most probable activating and deactivating interactions, respectively, selected
from Fig.~\ref{connectionfig}. The activating arrow for Cln3 is not a clear winner, therefore colored gray.
Nodes are placed in the order of their activation in the cell-cycle process, starting from Cln3 and continues clockwise,
SBF and MBF turn on at the same time, and similarly for Mcm1/SFF and Clb1,2.
Positive feed-forward interactions connecting Cln3 to Cln1,2, Cln3 to Clb5,6 and Clb5,6 to Mcm1/SFF can be seen.
Negative feedback connections from later nodes to earlier nodes are also obvious.
These interactions will be more common when more arrows are added.
This network (less the gray arrow) does not produce the cell-cycle process but one very close to it.}
\label{ccfeedbackfig}
\end{figure}

\begin{figure}
\includegraphics[width=\columnwidth]{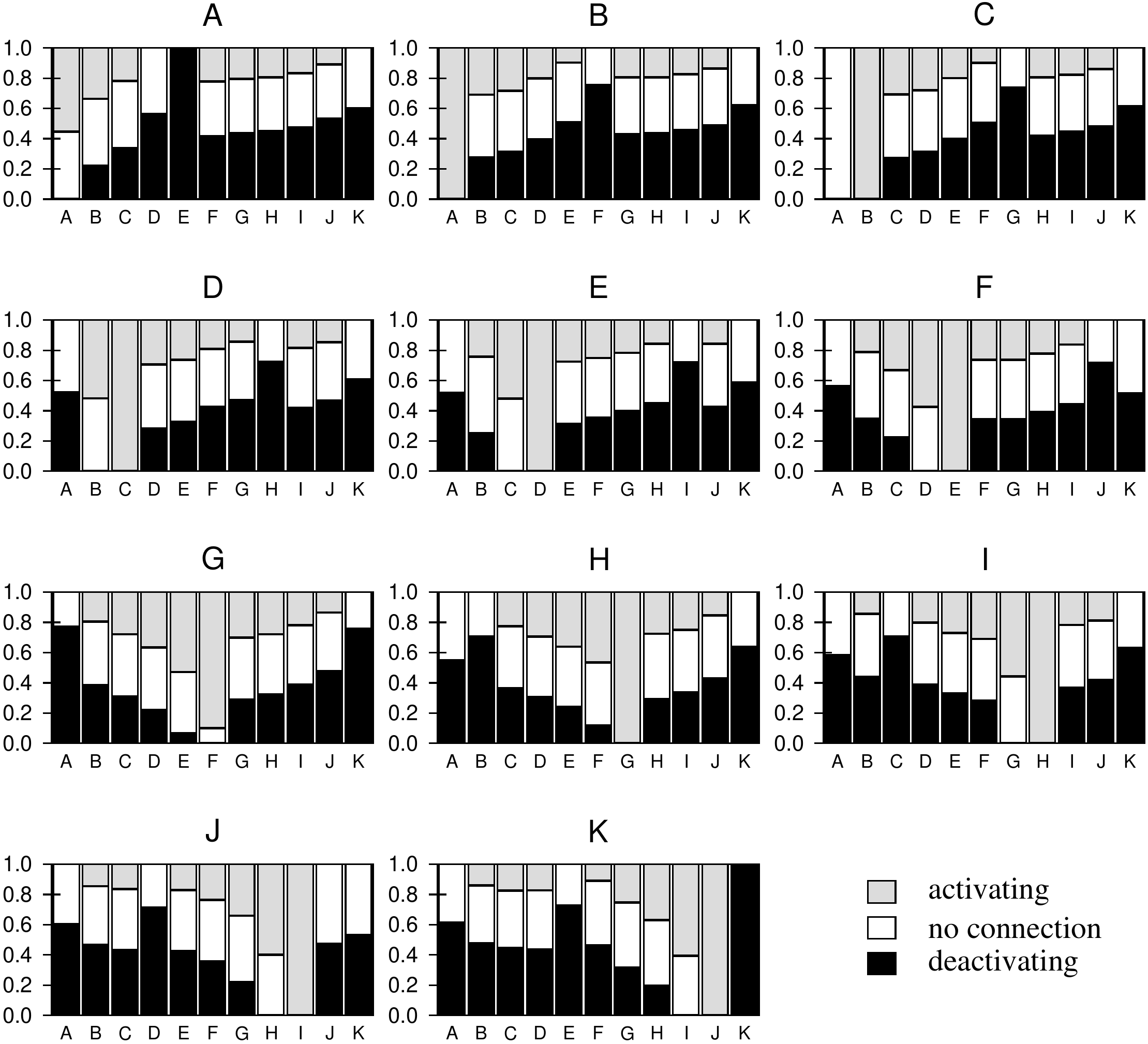}
\caption
{Fractions of different types of inward connections to each node from all other nodes (including itself) that produce the activation cascade in Table~\ref{cascade5tab}.}
\label{connection5cyclefig}
\end{figure}

From Fig.~\ref{connectionfig} we can deduce that there are
10 core connections (shown as black solid and black dashed arrows in
Fig.~\ref{networkfig}) that are absolutely required in order to
produce the cell-cycle process. These required connections become
obvious once we look closer into the cell-cycle process. For example,
comparing the $G_1$ stationary state and the ``excited'' $G_1$ state,
Cln3 is the only node that is turned on; this implies that MBF and
SBF, which are turned on in the next time step, can only be activated
by Cln3. The remaining required connections can all be deduced using
the same logic. The compositions of different connection types follow
a common trend where there is a higher chance for a node to be
positively regulated by nodes that are active earlier in the
cell-cycle process (positive feed-forward) and negatively regulated by
nodes that are active later in the process (negative feedback) [see Fig.~\ref{ccfeedbackfig}].
This trend seems to be general for networks that produce cascades of
activation. To check this, we looked at the connection compositions
for two simple activation cascades, where 11 nodes are activated in
turn for 4 or 5 time steps, and they indeed show the same trend [See
Table~\ref{cascade5tab} and Fig.~\ref{connection5cyclefig} for 5 time
steps activation cascade].

\begin{figure}
\includegraphics[width=\columnwidth]{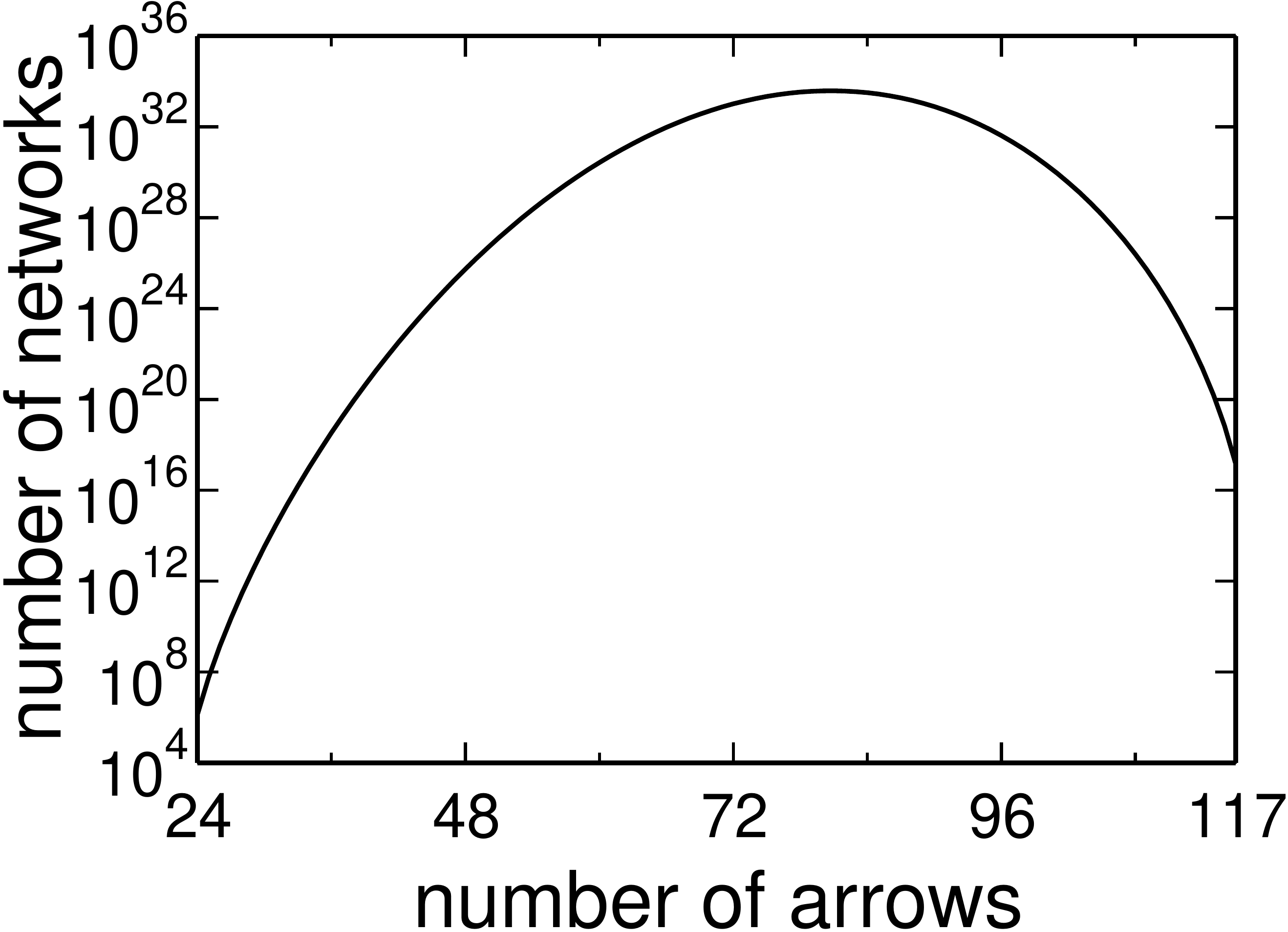}
\caption
{Distribution of number of networks that can realize the cell-cycle process over the number of arrows in the networks.}
\label{netcountvsacountfig}
\end{figure}
\begin{table}[ht]
\caption
{A simple activation cascade with 11 nodes activated in turn for 5 time steps}
\begin{tabular*}{\columnwidth}{@{\extracolsep{\fill}} cccccccccccc}
\hline
\hline
Time & A & B & C & E & F & G & H & I & J & K & L\\
\hline
1 & 1 & 0 & 0 & 0 & 0 & 0 & 0 & 0 & 0 & 0 & 0\\
2 & 1 & 1 & 0 & 0 & 0 & 0 & 0 & 0 & 0 & 0 & 0\\
3 & 1 & 1 & 1 & 0 & 0 & 0 & 0 & 0 & 0 & 0 & 0\\
4 & 1 & 1 & 1 & 1 & 0 & 0 & 0 & 0 & 0 & 0 & 0\\
5 & 1 & 1 & 1 & 1 & 1 & 0 & 0 & 0 & 0 & 0 & 0\\
6 & 0 & 1 & 1 & 1 & 1 & 1 & 0 & 0 & 0 & 0 & 0\\
7 & 0 & 0 & 1 & 1 & 1 & 1 & 1 & 0 & 0 & 0 & 0\\
8 & 0 & 0 & 0 & 1 & 1 & 1 & 1 & 1 & 0 & 0 & 0\\
9 & 0 & 0 & 0 & 0 & 1 & 1 & 1 & 1 & 1 & 0 & 0\\
10 & 0 & 0 & 0 & 0 & 0 & 1 & 1 & 1 & 1 & 1 & 0\\
11 & 0 & 0 & 0 & 0 & 0 & 0 & 1 & 1 & 1 & 1 & 1\\
12 & 0 & 0 & 0 & 0 & 0 & 0 & 0 & 1 & 1 & 1 & 1\\
13 & 0 & 0 & 0 & 0 & 0 & 0 & 0 & 0 & 1 & 1 & 1\\
14 & 0 & 0 & 0 & 0 & 0 & 0 & 0 & 0 & 0 & 1 & 1\\
15 & 0 & 0 & 0 & 0 & 0 & 0 & 0 & 0 & 0 & 0 & 1\\
16 & 0 & 0 & 0 & 0 & 0 & 0 & 0 & 0 & 0 & 0 & 0\\
\hline
\hline
\end{tabular*}
\label{cascade5tab}
\end{table}

Next, we generated two network ensembles for each number of arrows
allowed in the space of cell cycle networks. This number of arrows
varied from 24 to 117 as seen in Fig.~\ref{netcountvsacountfig}.  The
first ensemble is a combined structural/functional ensemble [see
Methods] consisting of 1,000 networks that both realize the cell cycle
and have a fixed number of arrows.  These networks will be referred to
as the ``cell-cycle networks'' (CN). The second ensemble was generated
by randomly reconnecting the arrows for each network in the first
ensemble [see Methods]. This ensemble will be referred to as the
``random networks'' (RN).

\begin{figure}
\includegraphics[width=\columnwidth]{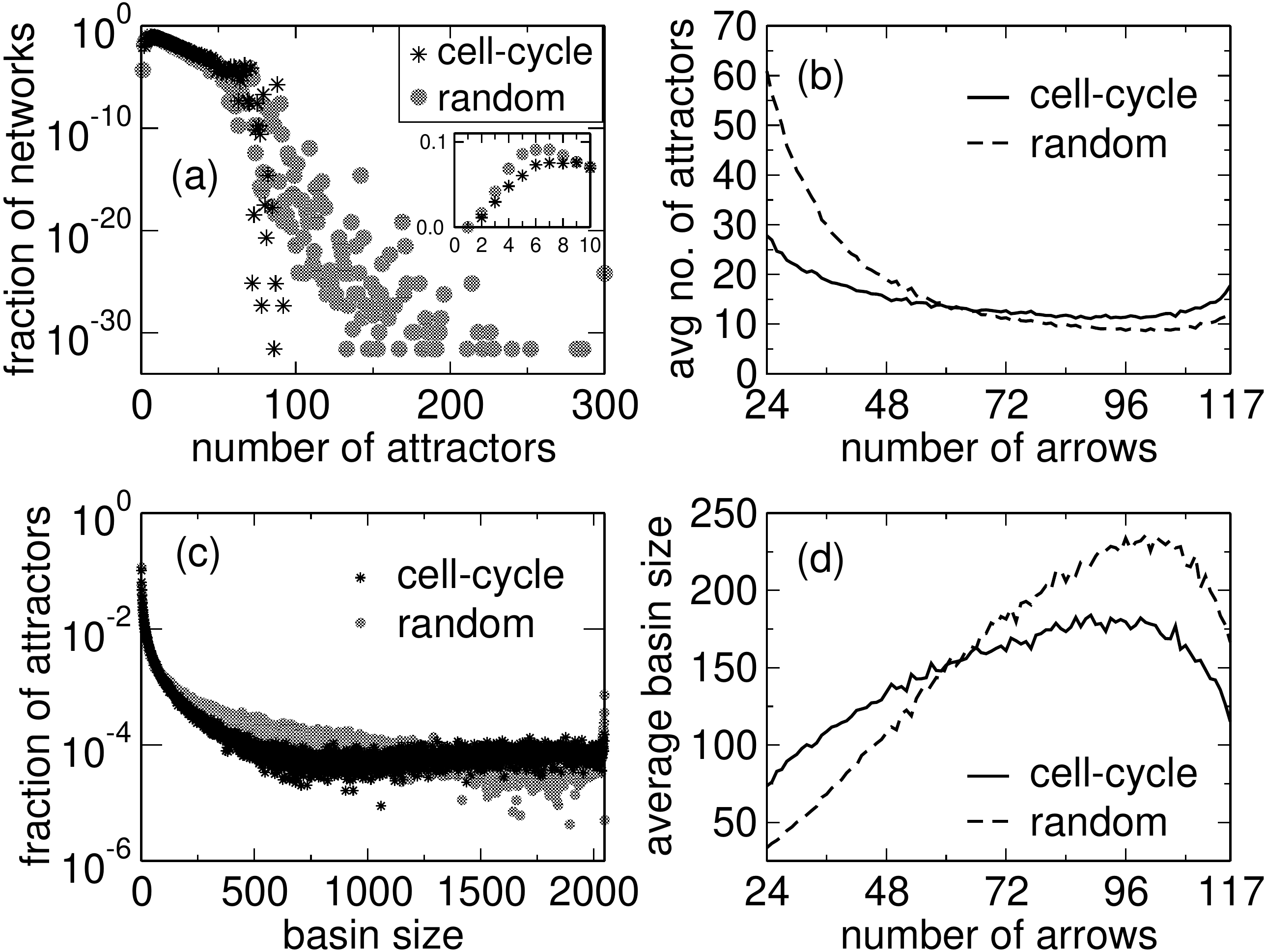}
\caption
{Number of attractors and basin sizes of attractors for the two ensembles of networks. (a) Distribution of number of attractors. Inset shows the distribution for $\leq$ 10 attractors (b) Number of attractors averaged over networks with the same number of arrows. (c) Distribution of size of basins of attractions. (d) Basin size averaged over networks with the same number of arrows.}
\label{numattsizefig}
\end{figure}

\subsection{Analysis of Attractors: Large Basins for Free}
We studied the time evolution of protein states of the two ensembles
by using the dynamical model described in the Methods and initiating
the networks from each of the $2^{11} = 2,048$ states. We found that
the CN networks have fewer attractors and larger attractor basin sizes
compared to the RN networks [see Fig.~\ref{numattsizefig} (a),(c)]. The number
of attractors decreases as the number of arrows increases, reaching a minimum at around 100 arrows
and then increases again [see Fig.~\ref{numattsizefig} (b)]. 
The two curves cross each other at about 60 arrows, therefore, compared to RN networks, CN networks
have fewer attractors when the number of arrows is small and more attractors when the number of arrows
is large. The behavior is reversed for the size of attractor basins [see Fig.~\ref{numattsizefig} (d)]. 
In the CN ensemble, the probabilities for a network with 34
arrows to have $\leq$ 7 attractors and to have the biggest attractor
basin size $\geq$ 1764 (as in the case of yeast cell-cycle network)
are 4.1\% and 7.5\% respectively. In the RN ensemble the
corressponding percentages are only 1.5\% and 0.7\% respectively.
Thus we see that the constraint of having to perform the yeast cell
cycle cascade alone can, to a certain extent, help explain the origins
of these two measures of dynamical robustness; a large basin
essentially arises for free as a consequence of the cell cycle
function.  Indeed, the average basin size of the biggest attractors for 
the CN ensemble is 1397 compared to 1045.53 for the RN ensemble. In addition, 
95.3\% of the networks in the CN ensemble have the
$G_1$ stationary state as the biggest attractor and the average basin
size of these attractors is 1536.92.
\begin{figure}
\includegraphics[width=\columnwidth]{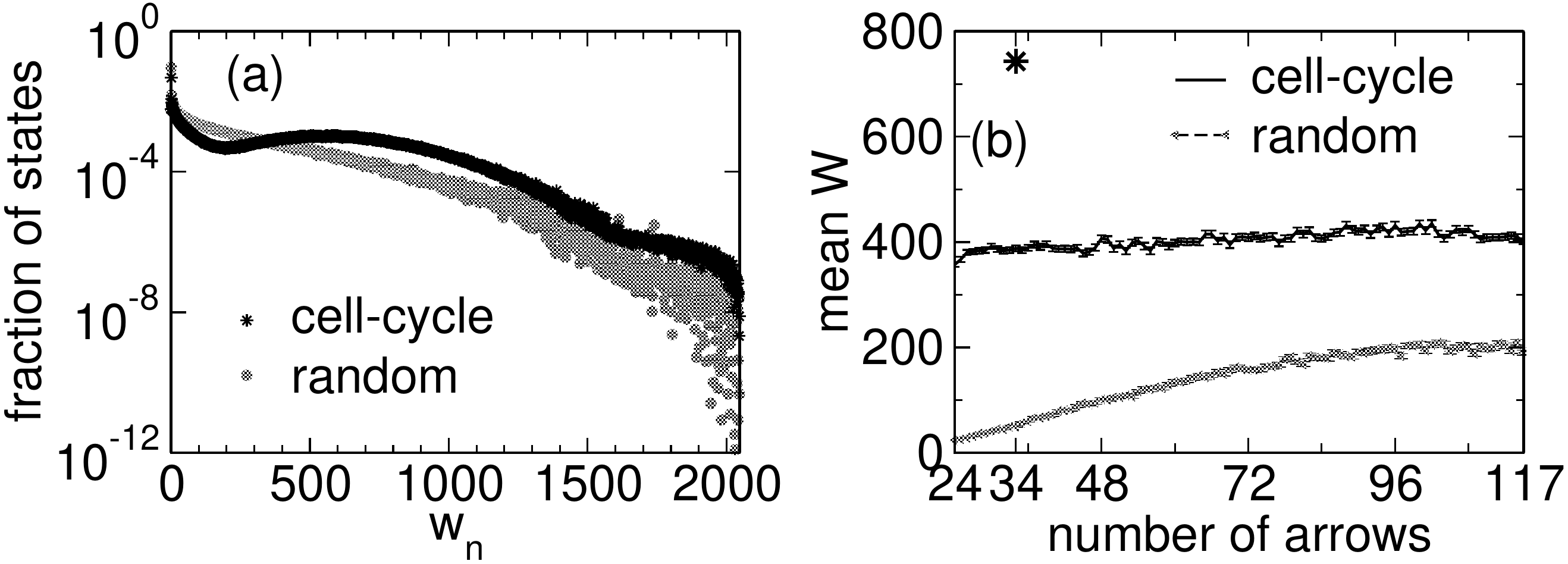}
\caption
{(a) Distribution of convergence value $w_n$ for network states in each ensemble. (b) Overall convergence $W$ averaged over networks with the same number of arrows. Error bar represent one standard error.}
\label{randoowaawfig}
\end{figure}
\subsection{Convergence of Trajectories.}
Following \cite{LiLonLuOuyTan}, we define a measure $w_n$ that
quantifies the ``degree of convergence'' of the dynamical network
trajectories onto each network state $n$ where $n=1,\dots,2048$.  Let
$T_{j,k}$ denote the number of trajectories starting from all 2048
initial network states that travel from state $j$ to state $k$ in one
time step.  Let $L_n$ denote the number of steps it takes to get from
state $n$ to its attractor, so that we can index the states along the
outward trajectory by $k=1,\dots,L_n$.  Then $w_n = \sum_{k=1}^{L_n}
T_{k-1,k}/L_n$.  The overall convergence, or overlap $W$ of trajectories
is given by the average of $w_n$ over all states $n$.
\par
The distribution of the $w_n$ values is shown in
Fig.~\ref{randoowaawfig} (a). The result shows that there are more states in
the CN ensemble having larger $w_n$ values indicating a higher degree
of convergence in the network dynamics. The local maxima at $w_n =
559$ for this ensemble should be a result of the requirement that
networks in this ensemble must produce the cell-cycle process. The
overall overlap $W$ [see Fig.~\ref{randoowaawfig} (b)], which is the average of $w_n$ over all network
states, for the yeast cell-cycle network is 743 and the probability
for a network with 34 arrows in the CN ensemble to have $W \geq 743$
is 3.1\%.  Such a result is highly unlikely in the RN ensemble.  In
fact no networks in the RN ensemble with 34 arrows had an overlap $W
\geq 743$.  Thus a higher degree of convergence is a dynamical
consequence of performing the cell cycle function, but nevertheless,
the actual cell cycle network in Fig.~\ref{networkfig} still displays
a relatively high degree of convergence even within the CN ensemble.  The
statistical significance of the yeast cell cycle's overlap parameter $W$
within the functional cell cycle ensemble suggests the existence of a
strong selection pressure for convergent dynamics.  

\begin{figure}
\includegraphics[width=\columnwidth]{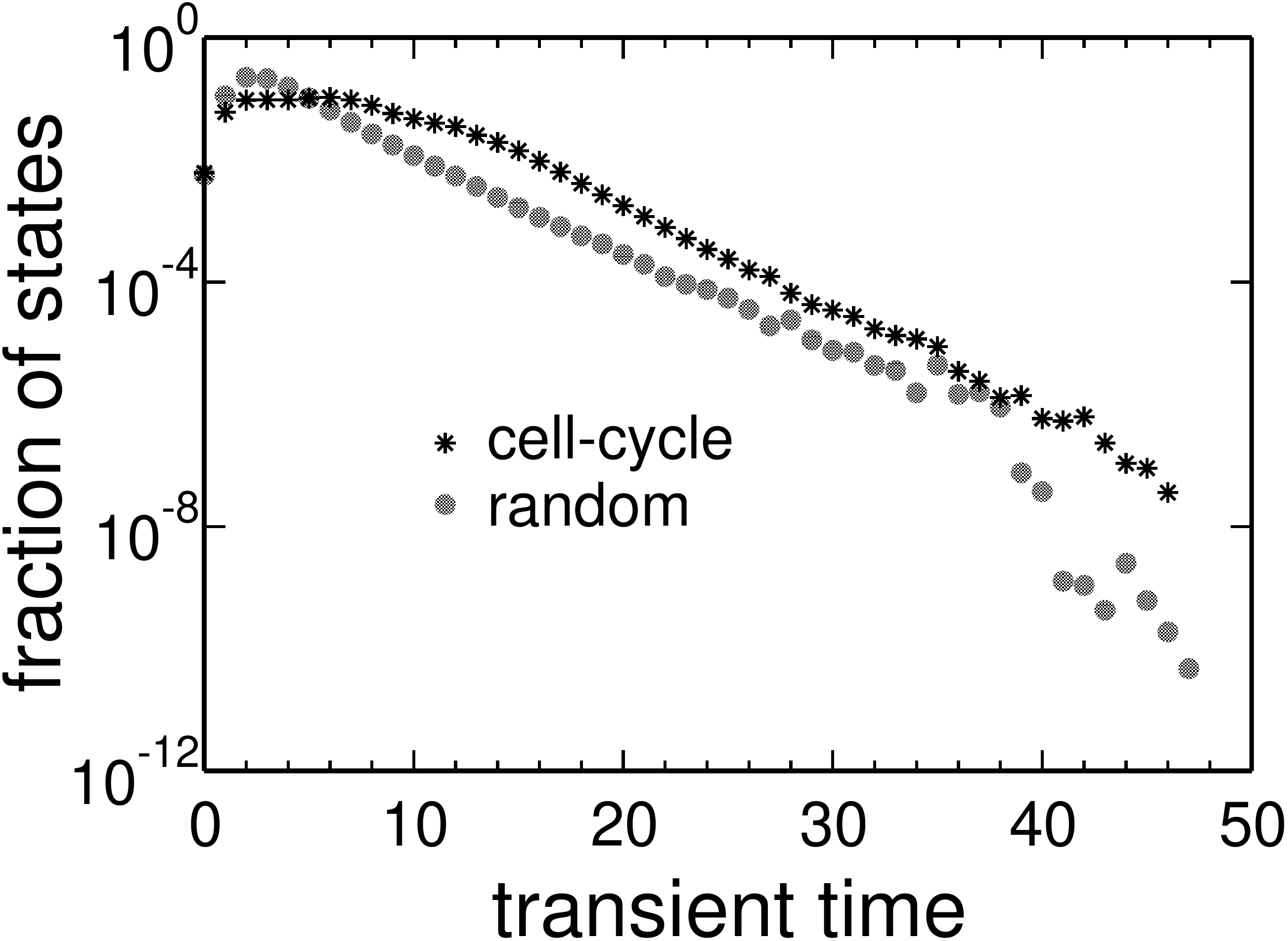}
\caption
{Distribution of transient times.}
\label{randoolpfig}
\end{figure}

\begin{figure}
\includegraphics[width=\columnwidth]{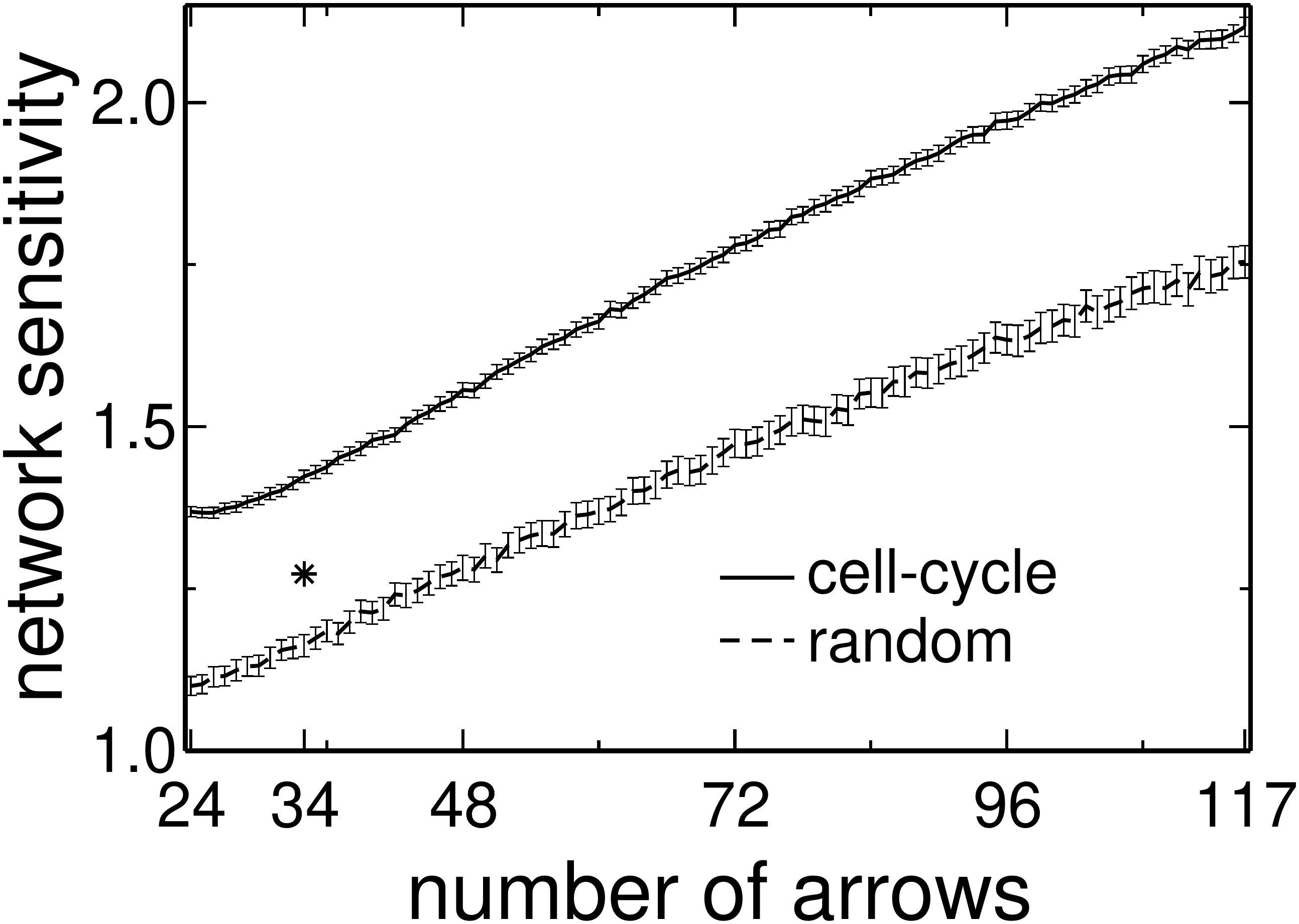}
\caption
{Network sensitivity averaged over networks with the same number of arrows. Asterisk ($\ast$) shows the network sensitivity of the yeast cell-cycle network. Error bar represent four times standard error.}
\label{randactivityavgfig}
\end{figure}
\subsection{Dynamical Order: Transients and Sensitivity.}
To compare the degree of order between the networks in the two
ensembles, we calculated the transient time for all network states for
all networks [see Fig.~\ref{randoolpfig}]. The transient time is defined
as the amount of time for a network state to evolve to its attractor,
which is equal to the length of its trajectory~\cite{Wuensche}. The
result shows that CN networks have longer transient times and thus are
more chaotic than RN networks (unless the RN networks have long
attracting limit cycles). The average transient time for the yeast
cell-cycle network is 7.47, and the probability for a cell-cycle
network with 34 arrows to have $\geq 7.47$ average transient time is
4.8\%.  
\par We then calculated the network sensitivity $s$~\cite{ShmuKau} for
all networks [see Fig.~\ref{randactivityavgfig}].  Network sensitivity
is the average expected number of node state changes in the output
given a one node state change in the input. In other words, $s$
calculates the average hamming distance of the output states of the
network for all hamming neighbor input states ({\it i.e.} hamming
distance = 1). If $s < 1$, the network is ordered; fluctuations in
node states die out quickly and remain localized. If $s > 1$, the
network is ``chaotic''; fluctuations are amplified at least on short
time scales.  On longer time scales, which are not captured,
especially for small networks, by this one step measure of
sensitivity, the network may not be chaotic and could still converge
to a stable fixed point.  When $s = 1$, the network is in some sense
critical.
\par The result in Fig.~\ref{randactivityavgfig} indicates that
networks in both ensembles are on average chaotic, or more precisely,
display a significant degree of short time scale sensitivity to input
perturbations, for any number of arrows within the range we
studied. The sensitivity $s$ increases monotonically with the number of arrows in a
network. The values of s for the RN ensemble remain smaller than those
for the CN ensemble demonstrating that CN networks are indeed more
chaotic, or sensitive to input perturbations, than RN networks.  The
yeast cell-cycle network has a network sensitivity of 1.27, which is
more ordered on average than other members in its ensemble.
Indeed the probability for a network with 34 arrows in the CN ensemble
to have $s \leq 1.27$ is only 3.1\%.  Thus the actual cell cycle is
remarkably ordered, or insensitive to input perturbations, despite the
fact that the functional constraint of performing the cell cycle
drives networks to be more sensitive to such perturbations on short
time scales.

\begin{figure*}
\includegraphics[width=\textwidth]{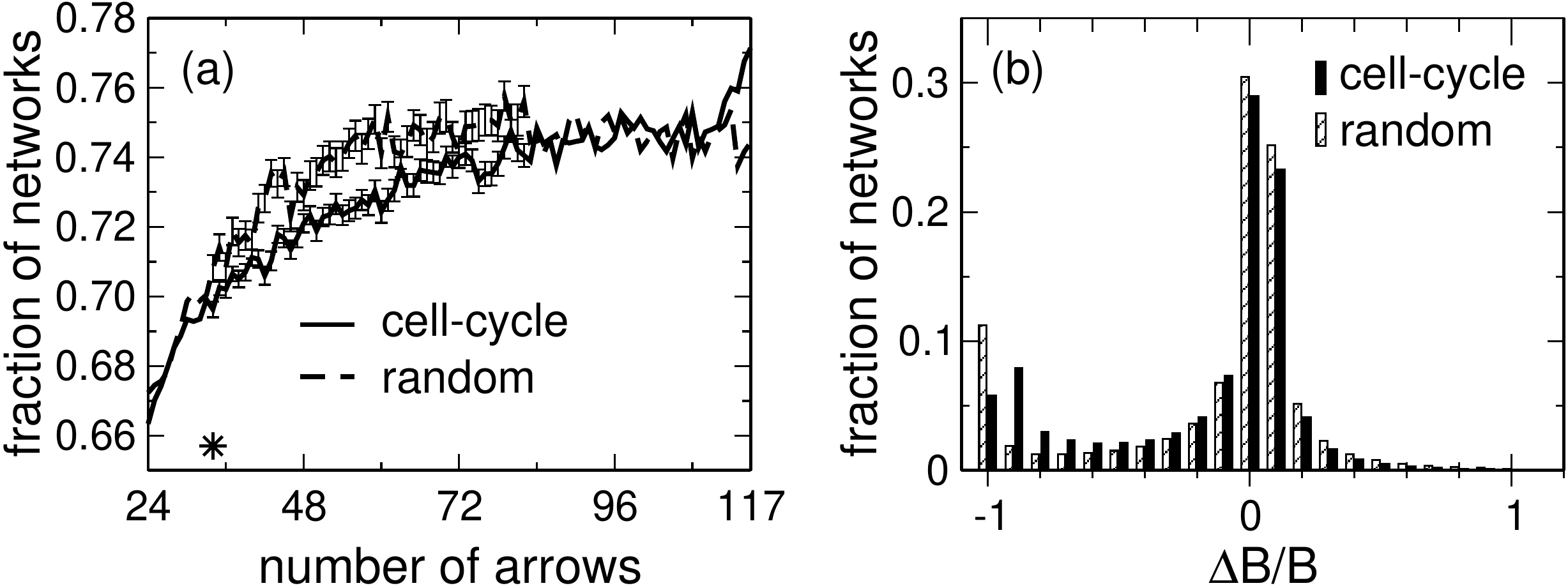}
\caption
{Response to perturbations to network structure. (a) Fractions of perturbed networks, for each number of arrows, that retain the original biggest attractor. Asterisk ($\ast$) shows the fraction for the perturbed yeast cell-cycle network. Error bars represent one standard error. (b) Distribution of relative changes of the size of the basin of attraction for the biggest attractor.}
\label{aucdbfig}
\end{figure*}
\subsection{Dynamical response to structural perturbations.}
To check and compare how the networks respond to structural
perturbations, we performed the same kinds of alterations described in
Li {et al.}~\cite{LiLonLuOuyTan} on all networks in the two
ensembles. The alterations included deleting arrows from, adding
arrows to and switching the signs of arrows in the networks. We
assessed the response by calculating the percentage of perturbed
networks that retain their original biggest attractor [see
Fig.~\ref{aucdbfig} (a)] and also the relative change in the size B of
the basin of attraction for the original biggest attractors [see
Fig.~\ref{aucdbfig} (b)]. The percentages of perturbed networks in the
two ensembles that retain their original biggest attractor both
increase initially when the number of arrows in the network is
small. The percentages are similar when there are about 24 to 33
arrows in the networks but as the number of arrows exceeds 33, the
percentage for the RN ensemble becomes significantly greater compared to that for
the CN ensemble. The two percentage become similar again when the
number of arrows increases beyond 83 and separate once more at 144 arrows but the
percentage for CN ensemble is greater this time.
The percentage for yeast cell-cycle
network (65.7\%) is smaller than the average for CN networks with the
same number of arrows. The probability to obtain an equal or higher
percentage is 80.7\%, indicating that the yeast cell-cycle network has
a worse than average robustness with respect to such structural
perturbations.
\par We noticed from the distributions of ${\Delta}B/B$ that there is
a higher chance for perturbations to have a deleterious effect to
networks in the CN ensemble where the change in the sizes of basins of
attraction is usually negative. However, there is a much higher chance
for networks in the RN ensemble to completely lose the original
biggest attractor (${\Delta}B/B = -1$), which is even more
unfavorable. The above effects should be attributed to the smaller
basins of attraction for networks in the RN ensemble. The average
${\Delta}B/B$ for yeast cell-cycle network is -0.342 and the
probability for a CN network with 34 arrows to have average
${\Delta}B/B$ value $\geq -0.342$ is 89\%. This again signifies a
worse than average robustness for the yeast cell-cycle network.

\begin{figure}
\includegraphics[width=\columnwidth]{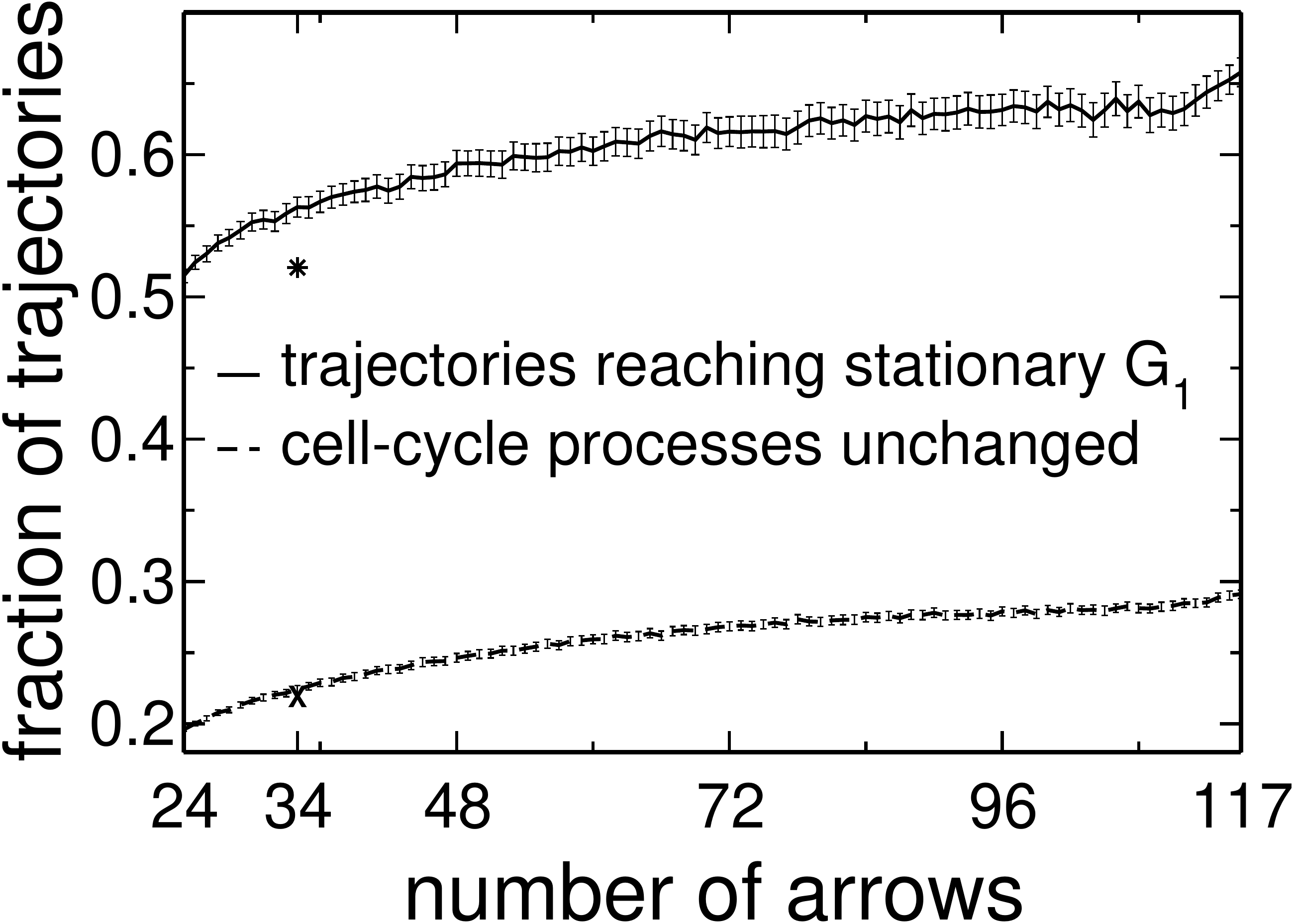}
\caption {Fractions of trajectories of the perturbed networks starting
  from the ``excited'' $G_1$ state that still evolve to the biological
  $G_1$ stationary state and fraction of cell-cycle processes of the
  perturbed networks remain unchanged.  Asterisk ($\ast$) and cross
  (x) show the fraction of trajectories reaching $G_1$ stationary
  state and fraction of cell-cycle processes unchanged, respectively,
  for perturbed yeast cell-cycle networks. Error bars reprensent four
  times standard error.}
\label{randtrajucfig}
\end{figure}
\subsection{Stability of the cell cycle process.}
Finally, we checked how many perturbed networks from the CN ensemble
could still maintain the cell-cycle process [see
Fig.~\ref{randtrajucfig}]. Starting from the ``excited'' $G_1$ state,
the fraction of trajectories reaching the $G_1$ stationary state and
the fraction of cell-cycle processes unchanged increase as the number
of arrows in the network increases. The fraction of trajectories
reaching the $G_1$ stationary state and the fraction of cell-cycle
processes unchanged for the yeast cell-cycle network are 0.52 and 0.22
respectively. The probability for a CN network with 34 arrows to
maintain $\geq 52$\% of trajectories reaching the $G_1$ stationary
state and $\geq 22$\% of cell-cycle processes unchanged are 79\% and
61.2\% respectively.

\section{Discussion}

We presented a maximum entropy analysis method that can reveal the
underlying structural constraints, as well as the statistical
significance of various dynamical properties, of networks that perform
a certain function.  We applied this method to the yeast cell-cycle
network and the accompanying cell-cycle process~\cite{LiLonLuOuyTan}.
\par
We demonstrated that requiring a network to produce an activation
cascade, {\it e.g.}  the cell-cycle process, requires the network to
have positive feed-forward and negative feedback interactions between
their nodes.  It is not just the case that this is a good design
principle to realize a long transient cascade; it is essentially the
only way to achieve it generically, and yields an example of how
network function constrains network structure.
\par
We also showed that certain dynamical features arise purely as a
consequence of performing the cell-cycle function.  Compared to the
random (RN) ensemble, networks in the cell-cycle (CN) ensemble had
much larger basins of attraction, a higher degree of convergence of
trajectories, longer transient times, and more chaotic behavior on
short time scales, as measured by the network sensitivity $s$.  These
properties may be essential for networks to produce a long sequence of
state transitions. The long trajectory of the cell-cycle process
provides many possible merge points for other trajectories, which
certainly contribute to the high degree of convergence in the network
dynamics and the large basin of attraction for $G_1$ stationary
state. Thus the existence of this globally attracting trajectory of
the dynamics alone can explain to a certain extent the observed
robustness against dynamical perturbations over long time scales.  
\par
On the other hand, with respect to structural perturbations, the
actual yeast cell cycle is relatively less robust compared to other
networks in the CN ensemble.  This is in stark contrast to the high
degree of dynamical order on short time scales displayed by the cell
cycle network, which suggests that there may be a trade off between
ordered dynamics and structural robustness. The network
sensitivity~\cite{ShmuKau}, which measures the degree of order,
calculates how sensitive a network is to fluctuations in the states of
the nodes, which is a major source of variation in a cell
population~\cite{EloLevSigSwa,RasOShea1,RasOShea2,SwaEloSig}. Evolution
may have favored a design for the yeast cell-cycle network that is
ordered and less sensitive to fluctuations in the states of the nodes
({\it e.g.} it has been reported that there is on average only 1 copy
of SWI5 mRNA per cell in yeast~\cite{Holland}), by sacrificing
robustness against perturbations to the network structure. However, we
expect that the complete yeast cell-cycle network is more robust
against such perturbations since it has ``redundant'' components and
connections that perform similar jobs.
\par
In any case, the observation that only 3.1\% of randomly chosen cell
cycle networks with 34 arrows are more ordered (as measured by the
sensitivity $s$) than the real cell cycle network reveals an
unsuspected but significant selection pressure for short time scale
ordered dynamics that cannot be explained by the functional
requirement of maintaining the cell cycle process; indeed the
functional requirement of maintaining the cell cycle proccess forces
the dynamics in the opposite direction, i.e. to be more chaotic on
short time scales.  Similarly, we have seen that simply requiring a
long cell cycle to occur automatically raises the average degree of
convergence of trajectories over longer time scales.  However, even
after accounting for this increase within the functional ensemble,
only 3.1\% of all cell cycle networks with 34 arrows have a greater
degree of convergence (as measured by the overlap parameter $W$),
reflecting an evolutionary pressure for convergent dynamics on long
time scales above and beyond that necessitated by the requirements of
the cell cycle function alone.
\par
Although we have focused on a single biological example, the cell
cycle, our analysis method is quite general. It would be interesting
to perform it on other mesoscopic scale networks that have a
comparable number of components to uncover their structural and
dynamical constraints.  More generally, we believe these techniques of
simultaneously analyzing both structural and functional ensembles of
networks will prove useful in the larger quest to deduce general
principles governing relations between structure, dynamics, function,
robustness and evolution.

\section{Acknowledgements}

We would like to thank Morten Kloster for insights into the sampling method.
C.T. acknowledges support from the Sandler Family Supporting Foundation and
 from the National Key Basic Research Project of China (2003CB715900). K.L. 
 acknowledges support from National Institutes of Health grant NIH GM067547.
S.G. acknowledges support from the Swartz foundation, and also thanks Hao Li
for useful discussions.  


\end{document}